\documentclass[reprint,superscriptaddress,aps,longbibliography]{revtex4-2}

\usepackage{comment}
\usepackage{mathtools}
\usepackage{physics}
\usepackage{amsmath,amssymb,amsfonts,amsthm}                                            
\usepackage{dcolumn}
\usepackage{bm}
\usepackage{enumerate}
\usepackage{hyperref}
\hypersetup{colorlinks,allcolors=blue} 
\usepackage{siunitx}
\usepackage{graphicx}

\graphicspath{{figs/}} 

\begin{document}

\title{Joint Unitarity and a Single Definite Outcome in a Quantum Measurement}

\author{Muxi Liu}
\affiliation{Institute for Quantum Studies, Chapman University, Orange, California 92866, USA}
\affiliation{Schmid College of Science and Technology, Chapman University, Orange, California 92866, USA}

\begin{abstract}
We investigated the possibility that a single measurement run with a definite outcome is a joint unitary evolution of all the participating systems, and measurement runs with different definite outcomes correspond to different unitary maps. With reasonable assumptions, we derived a lower bound of the dependence of the environment after measurement on the state of the system before measurement, conditioned on the same measurement outcome. An experimental test of this dependence relation can either serve as evidence that the unitary dynamics and the definite outcome in the orthodox sense cannot be true together or suggest a deviation from the von Neumann measurement model plus a ``conditioning" interpretational step.
\end{abstract}

\maketitle

\section{Introduction}
In the orthodox formulation of quantum mechanics, the global closed system always evolves unitarily, whereas the dynamics of a single run of projective measurement that generates a definite outcome is described by a special state‑update rule that differs in kind from all other dynamics. However, from an external physical standpoint, a projective measurement is essentially an interaction among systems---the system being measured, the system that records the outcome, and the relevant environment. Thus, it is natural to ask if the interaction-based overall unitary description can reproduce the same dynamics that the special state-update rule stipulates for a single measurement process.

The answer is usually taken to be no. The conventional way to model measurement as an interaction is the von Neumann model: the measurement coupling correlates the system’s eigenstates with pointer states of the apparatus. For inputs in superpositions, the unitary coupling produces an entangled system–apparatus state, so that neither reduced subsystem is left in an observable eigenstate. However, according to the eigenvalue-eigenstate link in the orthodox interpretation, eigenstates are the states in which systems have definite properties. Therefore, following this model, after a measurement, the apparatus does not have any definite record, and the system does not have any definite observed property. Many approaches have been developed to resolve this discrepancy. Most of them try to prevent the existence of such an entangled final state by modifying the unitary dynamics of quantum mechanics, for example with collapses, or try to explain the definite outcome in terms of such an entangled final state by changing the meaning of quantum states \cite{FamousOverall,FamousCollapse,MoreRecent}.

However, quantum mechanics works spectacularly well in every domain where it has ever been tested. In particular, in all non‑measurement scenarios in which the system or subsystem evolves into or asymptotically approaches a state associated with a definite eigenvalue of the relevant observable, the theory can describe these cases correctly. For example, in certain phases of a Rabi cycle when the system evolves to have probability 1 in the ground state or the excited state, the theory describes these time evolutions correctly; and in the final stage of cooling a subsystem to its ground state, the theory also describes correctly that the cooled subsystem asymptotically approaches the ground-state subspace. However, likewise, after a definite outcome has been generated after a single run of measurement, the system and the apparatus also have evolved to have probability 1 in their respective eigenstates or eigenspaces corresponding to the definite outcome that was just generated. This time, however, it appears that we have difficulties mentioned above to correctly describe these time evolutions. Is this interaction that we call the ``measurement" genuinely exceptional, demanding that we view quantum states differently or modify the fundamental law of dynamics? To answer this question, here we investigate the following possibility: orthodox quantum mechanics without the measurement state-update rule can correctly describe an individual measurement as a joint unitary evolution of the system, apparatus and environment that yields a definite outcome, in opposition to the von Neumann model of measurement.

Recent works effectively departing from the von Neumann measurement model have made efforts to model it differently to reproduce the predictions of quantum mechanics without postulating a collapse or changing the meaning of the states \cite{SuperdeterministicToyModel,MattSent,LeHu}. These works mainly focused on the dynamics of the system being measured and modeled it as either an open-system-type dynamics or a stochastic unitary evolution. They managed to reproduce the predictions of quantum mechanics using their respective choices of the specific form of the dynamics. However, these approaches are limited in the sense that they mainly focused on the measured system per se and treated the environment as a background rather than also a target of analysis and that their discussions mainly surrounded the specific models they respectively picked. In this work, we adopt a different strategy. Rather than proposing another detailed time‑resolved model, we ask a more prior question: if we only require that the overall dynamics of all the interacting ingredients is unitary in a measurement, what necessary conclusions can we derive with general reasonable assumptions and can we test them in experiments?

Our question will not make sense without the following crucial assumption, which is also the possible deviation from the von Neumann model that we are going to explore here: when we follow the apparently same operational steps to perform a projective measurement on an identically prepared system, the actual physical mechanism of the measurement process is not always the same. In particular, measurement processes that generate different outcomes from the same initial state of the system have different underlying mechanisms. This assumption is motivated by 
the possibility that in practice the measurement setup is generally not well controlled at the microscopic level. It is also this assumption that allows us to bypass the challenge that the linearity of quantum mechanics introduces to the standard von Neumann model. 

Furthermore, we assume that there is no special mechanism which demands that an underlying mechanism that generates a particular outcome can never occur when the initial joint input is in particular states. That is, we assume that an underlying mechanism that occurs when a particular outcome is generated from a particular joint input state is also possibly the one that occurs in any other possible cases with different initial states and hence needs to have the universal ability to generate the particular outcome from any possible joint input state. With this universality requirement, we expect that if we change the initial state of the system and keep the initial environment the same, then we will have a different environment after measurement if we obtain the same outcome. The reason is that unitary maps preserve information. Therefore, if a detector obtained a measurement record and ``collapsed" the system state, we should be able to subsequently find in the environment some trace of the system state before the ``collapse".

Our goal in this paper is to present our exploration of this possibility of the dynamical mechanism of quantum measurement, with our hope that it can provide different perspectives or insights on this topic. This paper is organized as follows: in Section \ref{sec 2}, we will first briefly review the von Neumann model and explain our multiplicity assumption of the underlying mechanisms of an operationally defined measurement in more concrete terms. Then, by imposing the initial and final joint states, we will derive a requirement on the initial environment for the dynamics to be possibly unitary. We will further impose the universality requirement on the underlying mechanism and derive a lower bound of the dependence of the final environment on the initial state of the system, conditioned on the same measurement outcome. In Section \ref{sec 3}, considering testing this dependence relation in an experiment, we will characterize two sources of noise that might blur the dependence signal that we can observe in an experiment and derive a noise-included lower bound on the dependence. In the discussion in Section \ref{sec 4}, we will clarify how we can test this dependence relation and the different implications of different potential results of the experimental test, with an explanation of how our proposed test can help us move forward with our understanding of quantum measurement. We will also reflect on and compare our approach and the question we ask with the conventional way of thinking about measurement and approaching the problem of outcomes. Finally, we will discuss possible related extensions in the future. A conclusion will be made in Section \ref{sec 5}. More detailed mathematical derivations are provided in Appendix \ref{appendix}.

\section{Can a measurement be globally unitary and generate a definite outcome?}\label{sec 2}
An important feature of measurement is that it reveals the definite property of a system when it is prepared to have a definite one. Therefore, in the case of measuring an n-level quantum system prepared in an eigenstate, the measurement should act as the map
\begin{equation}
    \ket{o_i}_S\ket{a_0}_A\ket{\phi_0}_E\xrightarrow[]{\hat{U}}\ket{o_i}_S\ket{a_i}_A\ket{\phi_i}_E.
\label{von Neumann}
\end{equation}
Here, $S$ denotes the system, $A$ the apparatus which is the single degree of freedom that registers the outcome, and $E$ the environment which includes all the other degrees of freedom that actually participate in this process. The map $\hat{U}$ is the unitary operator that describes the measurement process. The system state $\ket{o_i}_S$ with $i\in\{1,2,3,...,n\}$ is an eigenstate of the measured observable $\hat{O}$. The ready state of apparatus is $\ket{a_0}_A$, and the initial environment state is $\ket{\phi_0}_E$. After measurement, the apparatus state that records the outcome is $\ket{a_i}_A$, and the environment changes to $\ket{\phi_i}_E$ accordingly due to the potential interaction.

A direct consequence following Eq.(\ref{von Neumann}) is that when the system is prepared in a superposition, the map $\hat{U}$, forced by linearity, must act as
\begin{equation}
    \sum_{i=1}^{n}\alpha_i\ket{o_i}_S\ket{a_0}_A\ket{\phi_0}_E\xrightarrow[]{\hat{U}}\sum_{i=1}^{n}\alpha_i\ket{o_i}_S\ket{a_i}_A\ket{\phi_i}_E.
    \label{von Neumann superposition}
\end{equation}
In the final state, neither the system nor the apparatus is in an eigenstate of their respective observable. Hence, according to the orthodox interpretation, this model implies that the system still has no definite property and the apparatus has not obtained any definite outcome.

This way of modeling a measurement process, known as the von Neumann model, has been generally accepted when thinking of measurement. Its discrepancy from the empirical fact that measurements have definite outcomes is known as the problem of outcomes and has motivated different modifications of the unitary dynamical law of quantum mechanics and different views on the meaning of quantum states.

The problem of outcomes is inevitable if the von Neumann model is the correct model of measurement processes. However, it remains possible that even if in every measurement we measure the same system state, following the same operational procedure, the underlying microscopic dynamics that the interacting systems really experience are still different in the case that the outcome states are different. That is, it can be that in one case,
\begin{equation}
    \sum_{i=1}^{n}\alpha_i\ket{o_i}_S\ket{a_0}_A\ket{\phi_0}_E\xrightarrow[]{\hat{U}^{(1)}}\ket{o_1}_S\ket{a_1}_A\ket{\psi_1}_E,
\end{equation}
and in another case,
\begin{equation}
    \sum_{i=1}^{n}\alpha_i\ket{o_i}_S\ket{a_0}_A\ket{\phi_0}_E\xrightarrow[]{\hat{U}^{(2)}}\ket{o_2}_S\ket{a_2}_A\ket{\psi_2}_E,
\end{equation}
where $\ket{\psi_1}$ and $\ket{\psi_2}$ are final environment states, and $\hat{U}^{(1)}$ and $\hat{U}^{(2)}$ are different underlying unitary maps corresponding to the same macroscopic operational procedure we meant to implement in the lab. Unlike the conventional von Neumann model in which the same unitary map $\hat{U}$ in Eq.(\ref{von Neumann}) appears in all measurement processes, this different modeling, considering the multiplicity in the underlying mechanisms, allows the same initial joint state to end up in different final joint states because the mechanisms are different.

This multiplicity assumption opens up the possibility, but whether the map that connects the initial and final states of a single measurement process can really be unitary deserves more careful examination. To be general, here we model a generic case of a single run of projective measurement as having an initial joint state 
\begin{equation}
    \rho^{(i)}_{SAE}=\rho_S\otimes\ket{a_0}\bra{a_0}_A\otimes\rho^{(i)}_E,
    \label{Initial Joint}
\end{equation}
with $\bra{o_\theta}\rho_S\ket{o_\theta}>0$, and a final joint state 
\begin{equation}
    \rho^{(f)}_{SAE}=\rho^{(f)}_{SE}\otimes\ket{a_\theta}\bra{a_\theta}_A,
    \label{Final Joint}
\end{equation}
where $\theta$ is the outcome obtained, and 
\begin{equation}
    \bra{o_\theta}\text{tr}_{E}\left(\rho^{(f)}_{SE}\right)\ket{o_\theta}\geq1-\delta,
    \label{Final system closeness}
\end{equation}
with some small $\delta$. Here, we have considered all the systems that participate in this process from the beginning of the measurement to the end of the measurement when the apparatus has registered a definite outcome. We included all the degrees of freedom other than the system being measured, and the ``apparatus" degree of freedom that registers the outcome, in $E$, the environment. To be general and to preserve the joint state's rank, we allowed the actual reduced state of the system per se after measurement to be imperfect but very close to $\ket{o_\theta}_S$. However, the apparatus degree of freedom ends up exactly in an eigenstate corresponding to the definite property of having observed $\theta$, which is the same state it ends up in after measuring a system prepared perfectly in $\ket{o_\theta}_S$. This strictly corresponds to the empirical fact that a measurement produces a definite outcome. Here for simplicity, we have modeled the apparatus degree of freedom with pure states, and the fact or possibility that it has a degenerate eigenspace is discussed later in Section \ref{sec 4}.

We note that since a unitary map preserves eigenvalues, then for it to ever be possible to connect $\rho^{(i)}_{SAE}$ and $\rho^{(f)}_{SAE}$ with a unitary map, the initial environment has to satisfy the condition that there exists a rank $D$ projector $\hat{\Pi}_D$ such that
\begin{equation}
    \text{tr}\left((I-\hat{\Pi}_D)\rho^{(i)}_E\right)\leq\epsilon
    \label{Initial Environment Requirement}
\end{equation}
for some small $\epsilon$, and $D\leq d_E/d_S$, in which $d_E$ and $d_S$ are the dimensions of the system and environment Hilbert spaces. That is, since $\rho^{(f)}_{SAE}$ necessarily has at least $(d_S-1)d_E$ small eigenvalues, then to accommodate the most-demanding case when $\rho_S$ is a maximally mixed state, $\rho^{(i)}_{SAE}$ needs to have at least $d_S(d_E-D)\geq(d_S-1)d_E$ small eigenvalues. This requirement amounts to that the initial environment needs to have an effectively ``concentrated" occupation in its entire Hilbert space such that any occupation beyond the dominant $D$-subspace is small. Our analysis is only in a finite dimensional case, but a similar spirit can be extended to an infinite dimensional case. We note that this requirement is generally satisfied by an environment state in real life which is usually in a Hilbert space much larger than the system's space and is usually well concentrated with a long tail of small eigenvalues. Therefore, the possibility is open for a unitary map to connect the initial and final joint states of a single measurement process. However, it is not enough for a unitary map to connect only one pair of initial and final states of one run of measurement because this same mechanism might occur in other runs of measurement as well.

Suppose that we have a single run of measurement with an outcome $\theta$ which has an underlying unitary mechanism $\hat{U}^{(\theta)}$. Without any evidence to further constrain the condition for this mechanism to occur, such as that it is allowed to occur only once in a lifetime or only with a particular initial state, we assume that this mechanism $\hat{U}^{(\theta)}$ can occur in any measurement process when the outcome is $\theta$. Hence, it has the responsibility to map all possible initial states $\rho^{(i)}_{SAE}$ in the form of Eq.(\ref{Initial Joint}) with all admissible $\rho_S$ satisfying $\bra{o_\theta}\rho_S\ket{o_\theta}>0$ and $\rho^{(i)}_E$ satisfying Eq.(\ref{Initial Environment Requirement}) in terms of a common fixed projector $\hat{\Pi}_D$--for example, relative to the energy basis--to some corresponding final states $\rho^{(f)}_{SAE}$ in the form of Eq.(\ref{Final Joint}) with some corresponding $\rho^{(f)}_{SE}$. 

As a consequence of this universality requirement, $\hat{U}^{(\theta)}$ has to map every pure joint state in the dominant D-subspace to approximately a state with the system and apparatus in $\ket{o_\theta}_S\ket{a_\theta}_A$. That is, for any joint state 
\begin{equation}
\ket{\psi}\in\text{span}\{\ket{o_i}_S\otimes\ket{a_0}_A\otimes\ket{e_k}_E:k\leq D\},
\end{equation}
where $\{\ket{e_k}_E\}$ is the relevant environment basis, we have the output of $\hat{U}^{(\theta)}$ being $2\sqrt{\delta}$-close to its projection onto $\ket{o_\theta}_S\ket{a_\theta}_A$:
\begin{multline}
    \norm{\hat{U}^{(\theta)}\ket{\psi}\bra{\psi}\hat{U}^{{(\theta)}^\dagger}-\frac{\hat{\Pi}^{(\theta)}_{SA}\hat{U}^{(\theta)}\ket{\psi}\bra{\psi}\hat{U}^{{(\theta)}^\dagger}\hat{\Pi}^{(\theta)}_{SA}}{\text{tr}\left(\hat{\Pi}^{(\theta)}_{SA}\hat{U}^{(\theta)}\ket{\psi}\bra{\psi}\hat{U}^{{(\theta)}^\dagger}\right)}}_1\\
    \leq2\sqrt{\delta},
\end{multline}
where $\norm{\cdot}_1$ is the trace norm, and $\hat{\Pi}^{(\theta)}_{SA}=\ket{o_\theta}\bra{o_\theta}_S\otimes\ket{a_\theta}\bra{a_\theta}_A$. It is obtained by applying Eq.(\ref{Final Joint}), Eq.(\ref{Final system closeness}) and the gentle measurement lemma in the form of lemma 9.4.1 of \cite{Gentle_Measurement_Lemma}---which states that a measurement generating the highly likely outcome causes only a little disturbance to the original quantum state---to the pure state case. In particular, for a basis input, we have
\begin{equation}
\hat{U}^{(\theta)}\ket{o_i}_S\otimes\ket{a_0}_A\otimes\ket{e_k}_E\approx\ket{o_\theta}_S\otimes\ket{a_\theta}_A\otimes\ket{\Tilde{e}_{i,k}}_E,
\label{Basiswise Approximation}
\end{equation}
where we defined an environment state
\begin{equation}
    \ket{\Tilde{e}_{i,k}}_E=\frac{\bra{o_\theta}_S\bra{a_\theta}_A\hat{U}^{(\theta)}\ket{o_i}_S\otimes\ket{a_0}_A\otimes\ket{e_k}_E}{\norm{\bra{o_\theta}_S\bra{a_\theta}_A\hat{U}^{(\theta)}\ket{o_i}_S\otimes\ket{a_0}_A\otimes\ket{e_k}_E}}
\end{equation}
for all $i$ and $k$. Since the system-apparatus component of the output is approximately fixed, a remarkable consequence of Eq.(\ref{Basiswise Approximation}) is that for a superposed initial system state, we have, for example,
\begin{multline}
  \hat{U}^{(\theta)}\left(\alpha\ket{o_\theta}_S+\beta\ket{o_{\theta'}}_S\right)\otimes\ket{a_0}_A\otimes\ket{e_k}_E\approx
    \\   \ket{o_\theta}_S\otimes\ket{a_\theta}_A\otimes\left(\alpha\ket{\Tilde{e}_{\theta,k}}_E+\beta\ket{\Tilde{e}_{\theta',k}}_E\right),
    \label{Information Transfer}
\end{multline}
in which we note that the $\{\alpha,\beta\}$ information that is originally stored in the observable eigenbasis of the initial system is mostly transferred to some basis of the final environment after this run of measurement. The intuition behind this is that information must go somewhere rather than simply being destroyed, in a spirit similar to the no-hiding theorem \cite{No-hiding_theorem}.

In general, even if we do not know the relevant basis in the final environment to extract the $\{\alpha,\beta\}$ information, we should still expect a dependence of the reduced final environment state on the initial system state across different runs of measurement in which the same $\hat{U}^{(\theta)}$ occurs. That is, for two generic input states with the same environment state yet different system states, $\rho_{SAE}^{(i)}=\rho_S\otimes\ket{a_0}\bra{a_0}_A\otimes\rho_E^{(i)}$ and $\rho_{SAE}'^{(i)}=\rho_S'\otimes\ket{a_0}\bra{a_0}_A\otimes\rho_E^{(i)}$, the reduced final environment states should be correspondingly different:
\begin{equation}
    \norm{\rho_E^{(f)}-\rho_E'^{(f)}}_1\geq\norm{\rho_S-\rho_S'}_1-8\sqrt{\delta},
    \label{Baseline dependence}
\end{equation}
where $\rho_E^{(f)}=\text{tr}_{SA}(\hat{U}^{(\theta)}\rho^{(i)}_{SAE}\hat{U}^{{(\theta)}^\dagger})=\text{tr}_S(\rho_{SE}^{(f)})$ and likewise for $\rho'^{(f)}_E$. Its derivation is shown in Appendix \ref{appendix} by repeatedly applying triangle inequalities and the gentle measurement lemma. The intuition is again that since the system-apparatus component of the output is approximately fixed, characterized by a small $\delta$, most of the change made in the initial state of the system is forced to transmit to a dependent change in the reduced state of the final environment.

Therefore, a measurement process can be globally unitary, and the definite measurement outcome can be described by the definite eigenstates, if measurements yielding different outcomes do not share the same underlying mechanism. If that is the case, then ideally, with our universality assumption of the underlying mechanism $\hat{U}^{(\theta)}$, we should expect a difference in the reduced states of the final environment when we measure two different states of the system and obtain the same outcome $\theta$, given that the same underlying mechanism $\hat{U}^{(\theta)}$ occurs. In the next section, we will consider some non-ideality that might blur the dependence we can observe when testing it in an experiment: we might not control the initial environment well, and it might not be the same underlying mechanism $\hat{U}^{(\theta)}$ that occurs every time when we obtain the outcome $\theta$.

\section{Two Sources of Noise}\label{sec 3}
Previously, we have assumed that $\hat{U}^{(\theta)}$ can be the mechanism of any individual measurement process that generates the outcome $\theta$---the universality requirement---and therefore derived the dependence relation in Eq.(\ref{Baseline dependence}). However, it is not guaranteed that every time we obtain the outcome $\theta$, the underlying mechanism is always exactly the same. That is, for the same operational procedure we implement and the same outcome $\theta$ we obtain, there can still be a run-to-run fluctuation in the underlying global Hamiltonian, which, under the short time of a projective measurement, can result in a run-to-run fluctuation in the actual unitary map. Suppose that in an experiment we repeatedly measure the same and different system states multiple times and select the cases in which we obtain $\theta$ to analyze. Let us denote the collection of unitary maps that have actually occurred by $\{\hat{U}^{(\theta)}_l\}$, with each $\hat{U}^{(\theta)}_l$ having an independent frequency of occurrence $w_l$, where each $w_l\geq0$ and $\sum_lw_l=1$. We impose the same universality requirement on each $\hat{U}^{(\theta)}_l$ that each $\hat{U}^{(\theta)}_l$ can be the mechanism of any possible individual run of measurement that generates $\theta$ from any admissible initial state, and therefore all our previous analysis with $\hat{U}^{(\theta)}$ applies to each $\hat{U}^{(\theta)}_l$. Let us define an auxiliary reference $\hat{U}_0^{(\theta)}$ as the central element in $\{\hat{U}^{(\theta)}_l\}$ that minimizes the averaged distance from all other elements in the set, and suppose that we have
\begin{equation}
    \sum_lw_l\|\hat{U}^{(\theta)}_l-\hat{U}_0^{(\theta)}\|_{\diamondsuit}\leq\gamma,
    \label{Mechanism Fluctuation}
\end{equation}
where $\|\cdot\|_\diamondsuit$ is the diamond norm, the standard measure of difference between two quantum channels, which is the maximum trace distance between the outputs of the quantum channels $\hat{U}^{(\theta)}_l$ and $\hat{U}_0^{(\theta)}$ applied to the same input, maximized over all possible input states, and $\gamma$ characterizes how similar the actual unitary maps are on average. 

Moreover, besides a run-to-run fluctuation in the underlying mechanisms, we might also have a run-to-run fluctuation in the initial environment due to our limited control ability. Similarly, let us denote the collection of initial environment states that actually appear in our experiment by $\{\rho_E^{(m)}\}$, with each $\rho_E^{(m)}$ having an independent frequency of appearance $v_m$, where each $v_m\geq0$ and $\sum_mv_m=1$. Let us define an auxiliary reference $\rho_E^{(0)}$ as the central element in $\{\rho_E^{(m)}\}$ that minimizes the averaged distance from all other elements in the set, and suppose that we have
\begin{equation}
    \sum_m v_m\norm{\rho_E^{(m)}-\rho_E^{(0)}}_1\leq\eta,
    \label{Environment Fluctuation}
\end{equation}
where $\eta$ characterizes our control of the initial environment on average.

Therefore, with these two run-to-run fluctuations, the statistical conclusion we draw from the experiment actually reflects an averaged final environment state
\begin{equation}
    \bar{\rho}_E(\rho_S)=\sum_{l,m}w_l v_m \rho^{l,m}_E(\rho_S)
    \label{Averaged Final Environment}
\end{equation}
where we have used the assumption that $w_r$ and $v_m$ are independent and that our experiment consists of a sufficiently large number of repetitions. We also see the final reduced state of the environment for each individual run, $\rho^{l,m}_E$, as a function of the initial state of the system $\rho_S$, and the function is determined by the actual unitary map $\hat{U}^{(\theta)}_l$ and the actual initial state of the environment $\rho_E^{(m)}$:
\begin{equation}
    \rho^{l,m}_E(\rho_S)=\text{tr}_{SA}\left(\hat{U}^{(\theta)}_l\rho_S\otimes\ket{a_0}\bra{a_0}_A\otimes\rho_E^{(m)}\hat{U}^{{(\theta)}^\dagger}_l\right).
    \label{Individual Final Environment}
\end{equation}
Since $\hat{U}^{(\theta)}_l$ is close to $\hat{U}^{(\theta)}_0$ on average, and $\rho_E^{(m)}$ is close to $\rho_E^{(0)}$ on average, by applying the universality requirement and hence Eq.(\ref{Baseline dependence}) to $\hat{U}^{(\theta)}_0$ and $\rho_E^{(0)}$, we have the noise-included dependence relation for an experiment:
\begin{equation}
    \norm{\bar{\rho}_E\left(\rho_S\right)-\bar{\rho}_E\left(\rho'_S\right)}_1\geq\norm{\rho_S-\rho'_S}_1-8\sqrt{\delta}-2\left(\eta+\gamma\right),
    \label{Dependence Relation with Noise}
\end{equation}
which can be derived by combining Eqs.(\ref{Baseline dependence}--\ref{Individual Final Environment}) and applying triangle inequalities, as shown in Appendix \ref{appendix}.

Therefore, when we measure different system states and obtain the same measurement outcome, the difference in the system states should result in a difference in the environment states after measurement, with a distinguishability not less than the system distinguishability minus the noises. The distinguishability is bounded by Eq.(\ref{Dependence Relation with Noise}), but the ability to distinguish them is not guaranteed for any arbitrary observable. Furthermore, since we might only have access to some coarse-grained observables of the environment, the difference in environment caused by different system states might exceed the resolution. This can be the reason why this difference in the final environment is not some evident phenomenon that we have been easily observing all the time. However, if we can improve the resolution and increase the system's influence on the environment, then we can improve the chance of detecting the difference in the environment. For example, as an idealized intuitive illustration, if the underlying joint Hamiltonian is approximately time-independent, and if in two cases the system is prepared in the ground state and in nearly the excited state with a very large energy gap, then after measuring both cases to be in the ground state, the big difference in the expectation value of the system energy might result in a detectable difference in the energy expectation value of the environment.

\section{Discussion}\label{sec 4}
In the beginning, we imposed the orthodox quantum framework without the special state-update rule and asked the question whether an individual run of measurement can be described as a quantum process in which the final state corresponds to the single definite outcome that the measurement generates. More specifically, we asked whether if we consider all the systems that participate in this process, the joint initial and final states can be connected by a unitary map. We noted that it is possible if measurement processes that generate different outcomes correspond to different unitary maps and hence different mechanisms. Furthermore, we showed that if this possibility is actual, then, with our assumptions, we expect a dependence of the final environment state on the initial system state. However, we have been treating a measurement process as the target of analysis, without specifying the necessary and sufficient conditions---for example, what Hamiltonian we should construct in an experiment---to have a measurement process in the first place.

We are not trying to clarify the ambiguity of what counts as a measurement, but we know there exist measurement processes. For example, for any experimenter who wants to study a measurement process, they can always have a testing ground where they implement a measurement and obtain an outcome by themselves. When this happens, it is a measurement. However, it is not obvious that the ``measurement-like" process that we modeled necessarily waits until a conscious being comes into play to show up. That is, a definite readout might have already appeared after the interaction between the system and some detector that the experimenter used to implement the measurement. If that is the case, our modeling and analysis apply to this interaction between the system, environment, and this detector. In other words, our model and analysis apply to a ``measurement-like" process in which a definite outcome appears. According to our experience, a measurement process implemented by an experimenter is ``measurement-like", but some components or steps of the measurement might also be ``measurement-like", no matter if we are willing to call it a measurement by itself.

Therefore, to test the dependence relation described in Eq.(\ref{Dependence Relation with Noise}), we can do the following: we first pick a part of the measurement as the target of investigation and identify the system, the apparatus degree of freedom and the environment. We then run the measurement and read the outcome recorded in the apparatus degree of freedom, which should have been in a definite eigenstate before we read it, if a ``measurement-like" process occurred within the scope we picked. We repeat this a few times with the same and different initial states of the system and post-select the cases with the same outcome to analyze. 

If a dependence of the final environment on the initial system is not observed---that is, the same final environment for whatever initial system---then there are the following possibilities. (1) The unitary dynamics of closed systems and the orthodox interpretation together cannot explain the data, if we consider the scope we picked as sufficient to yield a definite outcome---for example, if a ``classical enough" detector and a wide enough range of environment are included, with an extreme having the experimenter themselves included. (2) The ``measurement-like" process occurs beyond the scope that we investigated. (3) There is a significant noise on average, which in the case that we approach nearly perfect control over the initial environment, indicates a significant averaged run-to-run difference in the underlying mechanisms such that it can arbitrarily bury any dependence signal. (4) The universality requirement does not hold. That is, some mechanisms are only responsible for mapping particular initial states to the outcome and never show up in other cases. (5) Even though in Eq.(\ref{Initial Joint}) and Eq.(\ref{Final Joint}) we considered everything other than the system and the single degree of freedom that records the outcome---the ``apparatus" degree of freedom---as part of the environment, there is still some hidden degeneracy associated with the ``apparatus" degree of freedom. This degeneracy is always large enough to hide the system information in the degenerate subspace, regardless of the dimension of the system's space. 

For possibilities (3), (4) and (5), without further assumptions and more specific modeling of the underlying unitary maps and the ``apparatus" degree of freedom, these possibilities will remain not testable and therefore irrelevant in terms of experimental verification. If we disregard options (3), (4) and (5), as the scope in a measurement process that we can investigate scales up, the chance of possibility (2) will be squeezed to 0, and it will appear more and more likely that option (1) is the case, indicating that the answer to our original question whether the dynamics of an individual measurement can be described as a globally unitary process in the orthodox framework is no.

However, if we observe, in any stage of a full measurement process, a dependence of the final environment on the initial system, then it suggests a deviation from the von Neumann model associated with a further ``conditioning" step of interpretation. It is also a signature of the unitarity of the individual measurement processes and a ``yes" to our original question. According to the conventional von Neumann model in Eq.(\ref{von Neumann superposition}), conditioning on a particular outcome, the environment always ends up in a fixed state regardless of the initial system state---as if every time the system ``actually" begins in that outcome state before measurement, and the measurement is only a process for the apparatus and the environment to note that. As a result, in terms of what we have access to in our lab before and after the measurement, the change of the joint quantum state is not unitary, and the initial information is lost. By contrast, if the dynamics in terms of what we have access to in our lab is unitary, like all other non-measurement processes that have been well tested, then, as we have shown, the information cannot be really lost. Therefore, even if the observed dependence is not a proof of unitarity, it is a signature.

The question we ask here is only on the dynamical level: whether the dynamics of an individual measurement is unitary. However, quantum measurements have not only a dynamical aspect regarding each individual run of measurement but also a statistical aspect regarding the statistics of the outcomes we obtain across different runs of measurement. The final state of the system in each individual run contains approximately no information of the initial system state because it has been approximately ``collapsed", whereas its appearance frequency across different runs of measurement of the same initial system state, conditioned on the same measurement basis, is ``statistically related" to the initial system state, which is an observational fact and is described by the Born rule.
If the dependence relation is observed and the individual dynamics might be unitary, then we might need to answer the further questions of what determines the specific mechanism and hence the outcome of an individual run of measurement, why its statistical distribution is in agreement with the Born rule, and why spacelike-separated measurements on entangled systems exhibit Bell-nonlocal outcome statistics. We do not yet have satisfactory answers to these questions.
It might be difficult to have satisfactory answers, and the answers might be testable or not. However, we want to stress that whether the dynamics of a single measurement, like that of all the non-measurement cases, is unitary in terms of what we have access to in our lab is not up to whether these associated questions are satisfactorily answered and is only up to the experiment. If the unitarity and the orthodox interpretation behind it are robust such that we observe the dependence relation, then the above ones are the difficult questions that we need to address the next. If the combination fails and we eventually do not observe the dependence relation, then we do not need to worry about these questions in the above sense but are left with a set of difficult dynamical and interpretational questions to address. In either case, concrete evidence regarding the dynamical issue of individual measurements can help us clarify the real set of questions we need to address in the future.

Conventionally, we have been mainly understanding the problem of outcome as a tension between the unitary dynamics and the interpretation, and this tension is already evident by conceptual and mathematical analysis alone. Perhaps this is because we have been thinking of measurement as a known quantum operation that we can precisely construct and implement. Therefore, the focus is mainly on the final state in Eq.(\ref{von Neumann superposition}), with either a different interpretation of it to resolve the tension or some additional laws of dynamics that prevent its existence to resolve the tension. However, in our work, we moved the focus to the measurement mechanism itself. Without assuming that we know the precise dynamical mechanism underlying a projective measurement, we hold fast on what we know about the initial and final joint states---within the orthodox interpretation of states---to have Eq.(\ref{Initial Joint}) and Eq.(\ref{Final Joint}) and wonder what the mechanism can be and, in particular, whether it can be globally unitary on all the participating systems. This shift of focus and question being asked makes it possible for the unitary dynamics and the orthodox interpretation to hold together in the individual runs of measurement, and this possibility can be tested in experiment.

Our work ultimately aims to promote our understanding of the mechanisms of quantum measurement. We hope that the different strategy we took and the possibility we explored can provide different insights or perspectives to our understanding of the dynamics of individual measurement processes. In the future, more comprehensive examinations of the alternative measurement model we explored, combined with the Born statistics that we observe, might help us better understand the more complete picture it might imply beyond the dynamics of an individual measurement process itself. In addition, an exploration of the potential explanations of Born statistics can be an important piece of a complete understanding of the mechanism of quantum measurement within this alternative modeling. Moreover, more specific modeling with further assumptions regarding the underlying unitary maps and the ``apparatus" degree of freedom in different concrete experimental scenarios can surely help us promote the experimental test. This will help us clarify the real set of questions that we need to address and advance our understanding of the mechanism of quantum measurement through concrete evidence.

\section{Conclusion}\label{sec 5}
We examined the possibility that an individual measurement process that generates a definite outcome can be described by a globally unitary map on all the participating systems within the orthodox interpretation due to a possible multiplicity in the underlying mechanisms of measurement processes that generate different outcomes. We examined the unitarity condition of individual measurement processes by imposing the initial and final joint states and obtained a requirement on the initial environment state. Then with a universality assumption and the characterizations of two potential sources of noise we might encounter in an experiment, we derived a lower bound of the dependence of the final environment state on the initial system state, in the cases of the same measurement outcome. We further clarified the experimental steps for testing the predicted dependence relation and the different implications of different potential experimental observations. In addition, we discussed potential questions related to the alternative measurement model that we explored here and stressed that an experimental test can tell us which set of difficult questions we will really need to address to understand quantum measurement. Moreover, we reflected on the strategy that we took and the question we asked and compared them with the conventional way of thinking about quantum measurement and approaching the problem of outcomes. We hope that our exploration can provide different perspectives or insights on this topic. Finally, we discussed potential directions in which we can move forward in the future.

\section{Acknowledgements}
I thank Professor Andrew N. Jordan for his guidance and support throughout this work. I thank Professor Matthew Leifer for helpful discussions and insightful feedback. I also thank Professor Ahmed Sebbar for helpful discussions. This work was supported by John Templeton Foundation under grant ID 63209.

\bibliography{main}

\appendix
\section{}\label{appendix}
To derive Eq.(\ref{Baseline dependence}), let us first define the auxiliary states
\begin{equation}
    \sigma_{SE}=\frac{\hat{\Pi}_S^{(\theta)}\rho^{(f)}_{SE}\hat{\Pi}_S^{(\theta)}}{\text{tr}\left(\hat{\Pi}_S^{(\theta)}\rho^{(f)}_{SE}\right)}
\end{equation}
and
\begin{equation}
    \sigma_{E}=\text{tr}_S\left(\sigma_{SE}\right)=\frac{\bra{o_\theta}_{S}\rho^{(f)}_{SE}\ket{o_\theta}_{S}}{\text{tr}\left(\hat{\Pi}_S^{(\theta)}\rho^{(f)}_{SE}\right)}.
\end{equation}
Then by Eq.(\ref{Final system closeness}) and the gentle measurement lemma in the form of lemma 9.4.1 of \cite{Gentle_Measurement_Lemma}, we have
\begin{equation}
    \norm{\rho^{(f)}_{SE}-\sigma_{SE}}_1\leq2\sqrt{\delta}.
    \label{Appendix 2}
\end{equation}
Further, by the contractivity of partial trace, we have 
\begin{align}
    \norm{\rho^{(f)}_{E}-\sigma_{E}}_1&=\norm{\text{tr}_{S}\left(\rho^{(f)}_{SE}\right)-\text{tr}_{S}\left(\sigma_{SE}\right)}_1\nonumber\\
    &\leq\norm{\rho^{(f)}_{SE}-\sigma_{SE}}_1\nonumber\\
    &\leq2\sqrt{\delta}.
    \label{appendix 1}
\end{align}
The same reasoning applies to the case of $\rho'^{(f)}_{E}$ and $\sigma'_{E}$. Therefore, we have
\begin{align}
    \norm{\rho^{(f)}_{E}-\rho'^{(f)}_{E}}_1&\geq\norm{\sigma_{E}-\sigma'_{E}}_1-\norm{\rho^{(f)}_{E}-\sigma_{E}}_1\nonumber\\
    &\quad-\norm{\rho'^{(f)}_{E}-\sigma'_{E}}_1\nonumber\\
    &\geq\norm{\sigma_{E}-\sigma'_{E}}_1-4\sqrt{\delta},
    \label{appendix 3}
\end{align}
where in the first line we have applied the triangle inequality and used the fact that $\rho^{(f)}_{E}-\rho'^{(f)}_{E}=(\sigma_{E}-\sigma'_{E})+(\rho^{(f)}_{E}-\sigma_{E})-(\rho'^{(f)}_{E}-\sigma'_{E})$, and in the second line we have applied Eq.(\ref{appendix 1}) twice for both cases. 

Moreover, since $\sigma_{SE}=\ket{o_\theta}\bra{o_\theta}_S\otimes\sigma_{E}$, we have
\begin{align}
    \norm{\sigma_{E}-\sigma'_{E}}_1&=\norm{\sigma_{SE}-\sigma'_{SE}}_1\nonumber\\
    &\geq\norm{\rho^{(f)}_{SE}-\rho'^{(f)}_{SE}}_1-\norm{\rho^{(f)}_{SE}-\sigma_{SE}}_1\nonumber\\
    &\quad-\norm{\rho'^{(f)}_{SE}-\sigma'_{SE}}_1\nonumber\\
    &\geq\norm{\rho^{(f)}_{SE}-\rho'^{(f)}_{SE}}_1-4\sqrt{\delta},
    \label{appendix 4}
\end{align}
where in the second line we have applied the triangle inequality and used the fact that $\sigma_{SE}-\sigma'_{SE}=(\rho^{(f)}_{SE}-\rho'^{(f)}_{SE})-(\rho^{(f)}_{SE}-\sigma_{SE})+(\rho'^{(f)}_{SE}-\sigma'_{SE})$, and in the third line we have used Eq.(\ref{Appendix 2}) twice for both cases.

Note that by Eq.(\ref{Initial Joint}) and Eq.(\ref{Final Joint}), we have
\begin{align}
    \rho^{(f)}_{SE}\otimes\ket{a_\theta}\bra{a_\theta}_A&=\rho^{(f)}_{SAE}=\hat{U}^{(\theta)}\rho^{(i)}_{SAE}\hat{U}^{(\theta)^\dagger}\nonumber\\
    &=\hat{U}^{(\theta)}\rho_S\otimes\ket{a_0}\bra{a_0}_A\otimes\rho_E^{(i)}\hat{U}^{(\theta)^\dagger}.
\end{align}
Since the initial environment component, $\rho^{(i)}_E$, is the same for both cases, and the unitary map preserves the trace distance, we have
\begin{equation}
    \norm{\rho^{(f)}_{SE}-\rho'^{(f)}_{SE}}_1=\norm{\rho_S-\rho'_S}_1.
    \label{appendix 5}
\end{equation}

Therefore, combining Eq.(\ref{appendix 3}), Eq.(\ref{appendix 4}) and Eq.(\ref{appendix 5}), we get
\begin{equation}
    \norm{\rho^{(f)}_{E}-\rho'^{(f)}_{E}}_1\geq\norm{\rho_S-\rho'_S}_1-8\sqrt{\delta},
\end{equation}
as in Eq.(\ref{Baseline dependence}).

To derive Eq.(\ref{Dependence Relation with Noise}), we first apply the triangle inequality to get
\begin{align}
    \norm{\bar{\rho}_E\left(\rho_S\right)-\bar{\rho}_E\left(\rho'_S\right)}_1
    &\geq\norm{\rho^{0,0}_E\left(\rho_S\right)-\rho^{0,0}_E\left(\rho'_S\right)}_1\nonumber\\
    &\quad-\norm{\bar{\rho}_E\left(\rho_S\right)-\rho^{0,0}_E\left(\rho_S\right)}_1\nonumber\\
    &\quad-\norm{\bar{\rho}_E\left(\rho'_S\right)-\rho^{0,0}_E\left(\rho'_S\right)}_1,
    \label{appendix 10}
\end{align}
where we have adopted the notation and definition in Eq.(\ref{Individual Final Environment}) and used the fact that 
\begin{align}
    \bar{\rho}_E\left(\rho_S\right)-\bar{\rho}_E\left(\rho'_S\right)&=\left(\rho^{0,0}_E\left(\rho_S\right)-\rho^{0,0}_E\left(\rho'_S\right)\right)\nonumber\\
    &\quad+\left(\bar{\rho}_E\left(\rho_S\right)-\rho^{0,0}_E\left(\rho_S\right)\right)\nonumber\\
    &\quad-\left(\bar{\rho}_E\left(\rho'_S\right)-\rho^{0,0}_E\left(\rho'_S\right)\right).
\end{align}
To proceed, we note that
\begin{align}
    \norm{\bar{\rho}_E-\rho^{0,0}_E}_1
    &=\norm{\sum_{l,m}w_lv_m\left(\rho^{l,m}_E-\rho^{0,0}_E\right)}_1\nonumber\\
    &\leq\sum_{l,m}w_lv_m\norm{\rho^{l,m}_E-\rho^{0,0}_E}_1\nonumber\\
    &\leq\sum_{l,m}w_lv_m\left(\norm{\rho^{l,m}_E-\rho^{l,0}_E}_1\right.\nonumber\\
    &\quad\left.+\norm{\rho^{0,0}_E-\rho^{l,0}_E}_1\right)
    \label{appendix 6}
\end{align}
for both $\rho_S$ and $\rho'_S$. We have used Eq.(\ref{Averaged Final Environment}) in the first line, the convexity of the trace norm in the second line, and the triangle inequality with the fact that $\rho^{l,m}_E-\rho^{0,0}_E=(\rho^{l,m}_E-\rho^{l,0}_E)-(\rho^{0,0}_E-\rho^{l,0}_E)$ in the third line. Furthermore, recalling the definition in Eq.(\ref{Individual Final Environment}) and the contractivity of partial trace, we have
\begin{align}
    \norm{\rho^{l,m}_E\left(\rho_S\right)-\rho^{l,0}_E\left(\rho_S\right)}_1&\leq\left\|\hat{U}^{(\theta)}_l\rho_S\otimes\ket{a_0}\bra{a_0}_A\right.\nonumber\\
    &\quad\left.\otimes\left(\rho_E^{(m)}-\rho_E^{(0)}\right)\hat{U}^{{(\theta)}^\dagger}_l\right\|_1\nonumber\\
    &=\norm{\rho_E^{(m)}-\rho_E^{(0)}}_1,
    \label{appendix 7}
\end{align}
where the last line is obtained because the unitary map preserves the trace distance. Similarly, we have 
\begin{align}
    \norm{\rho^{0,0}_E\left(\rho_S\right)-\rho^{l,0}_E\left(\rho_S\right)}_1&\leq\left\|\hat{U}^{(\theta)}_l\rho_S\otimes\ket{a_0}\bra{a_0}_A\right.\nonumber\\
    &\left.\quad\quad\quad\otimes\rho_E^{(0)}\hat{U}^{{(\theta)}^\dagger}_l\right.\nonumber\\
    &\left.\quad-\hat{U}^{(\theta)}_0\rho_S\otimes\ket{a_0}\bra{a_0}_A\right.\nonumber\\
    &\left.\quad\quad\quad\otimes\rho_E^{(0)}\hat{U}^{{(\theta)}^\dagger}_0\right\|_1\nonumber\\
    &\leq\norm{\hat{U}^{(\theta)}_l-\hat{U}^{(\theta)}_0}_\diamondsuit,
    \label{appendix 8}
\end{align}
where in the last line we have used the definition of the diamond distance. Therefore, since the same reasoning applies to the case of $\rho'_S$, by inserting Eq.(\ref{appendix 7}) and Eq.(\ref{appendix 8}) into Eq.(\ref{appendix 6}), we obtain
\begin{align}
    \norm{\bar{\rho}_E-\rho^{0,0}_E}_1
    &\leq\sum_{l,m}w_lv_m\norm{\rho_E^{(m)}-\rho_E^{(0)}}_1\nonumber\\
    &\quad+\sum_{l,m}w_lv_m\norm{\hat{U}^{(\theta)}_l-\hat{U}^{(\theta)}_0}_\diamondsuit\nonumber\\
    &\leq\eta+\gamma
    \label{appendix 9}
\end{align}
for both cases of $\rho_S$ and $\rho'_S$. In the last line, we have used Eq.(\ref{Mechanism Fluctuation}), Eq.(\ref{Environment Fluctuation}) and the fact that $\sum_lw_l=1=\sum_mv_m$.

Inserting Eq.(\ref{appendix 9}) back into Eq.(\ref{appendix 10}), we obtain 
\begin{align}
    \norm{\bar{\rho}_E\left(\rho_S\right)-\bar{\rho}_E\left(\rho'_S\right)}_1&\geq\norm{\rho^{0,0}_E\left(\rho_S\right)-\rho^{0,0}_E\left(\rho'_S\right)}_1\nonumber\\
    &\quad-2\left(\eta+\gamma\right).
    \label{appendix 12}
\end{align}
Further, by applying the universality requirement and hence the general dependence relation in Eq.(\ref{Baseline dependence}) to the specific case with $\hat{U}^{(\theta)}_0$ and $\rho^{(0)}_E$, we have
\begin{equation}
    \norm{\rho^{0,0}_E\left(\rho_S\right)-\rho^{0,0}_E\left(\rho'_S\right)}_1\geq\norm{\rho_S-\rho'_S}_1-8\sqrt{\delta}.
    \label{appendix 11}
\end{equation}

Therefore, by inserting Eq.(\ref{appendix 11}) into Eq.(\ref{appendix 12}), we finally obtain
\begin{equation}
    \norm{\bar{\rho}_E\left(\rho_S\right)-\bar{\rho}_E\left(\rho'_S\right)}_1\geq\norm{\rho_S-\rho'_S}_1-8\sqrt{\delta}-2\left(\eta+\gamma\right),
\end{equation}
as in Eq.(\ref{Dependence Relation with Noise}).

\end{document}